\documentclass[ aps, showpacs, showkeys, nofootinbib, floatfix, superscriptaddress]{revtex4}

\usepackage{amsfonts}

\usepackage{amssymb}
\usepackage{amsmath}
\usepackage{graphicx}

\begin{document}

\title{Cosmic acceleration and the change of the Hubble parameter}

 \author{G. Y. Chee}
 \email{qgy8475@sina.com}
 \affiliation{College of Physics and Electronics, Liaoning Normal University, Dalian,\\
 116029, China}

\begin{abstract}
A new model of accelerating expansion of the universe is presented.
A universal vacuum de Sitter solution is obtained. A new explanation
of the acceleration of the cosmic expansion is given. It is proved
that the changing of the expansion from decelerating to accelerating
is an intrinsic property of the universe without need of an exotic
dark energy. The cosmological constant problem, the coincidence
problem and the problem of phantom divide line crossing are
naturally solved.

\end{abstract}



\maketitle

\section{$\text{Introduction}$}

Very recently, there have been proposals for constructing generalizations of
teleparallel gravity in Refs. \cite{1}, \cite{2} and \cite{3} which followed
the spirit of $f(R)$ gravity (see \cite{4} for a review) as a generalization
of general relativity. That is, the Lagrangians of the theories were
generalized to the form $f(T)$, where $f$ is some suitably differentiable
function and $T$ is the Lagrangian of teleparallel gravity \cite{5}. The
interest in these theories was aroused by the claim that their dynamics
differ from those of general relativity but their equations are still second
order in derivatives and, therefore, they might be able to account for the
accelerated expansion of the universe and remain free of pathologies. It has
been shown, however, that this last expectation was unfounded: these
theories are not locally Lorentz invariant and appear to harbour extra
degrees of freedom \cite{6}. Even if one decides to give up teleparallelism,
such actions would not make sense as descriptions of the dynamics of gravity
if local Lorentz symmetry was restored \cite{7}. Based on the definition
given in these theories, the Weitzenbock connection, the torsion tensor and
the contortion tensor are not local Lorentz\ scalars (i.e. they do not
remain invariant under a local Lorentz transformation in tangent space).
This is the root of the lack of Lorentz invariance in generalized
teleparallel theories of gravity. The $f(T)$ models behave very differently
from the $\Lambda $CDM model on large scales, and are, therefore, very
unlikely to be suitable alternatives to it \cite{8}. Since this theory is
not invariant under local Lorentz transformations, the choice of tetrad
plays a crucial role in such models. Different tetrads will lead to
different field equations which in turn have different solutions \cite{9}.

It is known that the Lagrangian $T$ of teleparallel gravity only differs
from the Riemann curvature scalar $R$ by a boundary term\cite{7,8}. Recall
that in general relativity the Einstein field equation can be derived from a
{\em reduced} Lagrangian $\Gamma =\left( \left\{ _\sigma {}^\mu {}_\nu
\right\} \left\{ _\rho {}^\nu {}_\mu \right\} -\left\{ _\mu {}^\nu {}_\nu
\right\} \left\{ _\sigma {}^\mu {}_\rho \right\} \right) g^{\rho \sigma }$
which only differs from the Riemann curvature scalar $R$ of the Christoffel
symbol $\left\{ _\sigma {}^\mu {}_\nu \right\} $ by a boundary term $\nabla
_\mu w^\mu $ with $w^\mu =g^{\nu \lambda }\left\{ _\nu {}^\mu {}_\lambda
\right\} -g^{\nu \mu }\left\{ _\nu {}^\lambda {}_\lambda \right\} $ \cite{10}%
. Following the spirit of $f(R)$ gravity we can propose a modified theory of
gravity called {\bf reduced }$f(R)${\bf \ gravity} or $f\left( \Gamma
\right) ${\bf \ gravity} given by a Lagrangian which is a function of{\rm \ }%
$\Gamma ${\rm : }${\cal L}_G=f\left( \Gamma \right) $. It will be seen in
this letter that this theory is identified as a metrical formulation of $%
f\left( T\right) $ gravity. In contrast with $f\left( T\right) $ gravity,
the new theory, $f\left( \Gamma \right) $ gravity respects local Lorentz
symmetry and harbour no extra degrees of freedom, since the dynamical
variable is the metric instead of the tetrad. At the same time it has the
advantage over $f(R)$ gravity that its field equations are second-order
instead of fourth-order and then is free of pathologies. The disadvantage of
$f\left( \Gamma \right) $ gravity is that its field equations do not respect
general covariance. The significant pay-off that will be suggested in this
letter is that $f(\Gamma )$ gravity may provide an satisfactory alternative
to conventional dark energy in general relativistic cosmology and a new
explanation of the acceleration of the cosmic expansion. The changing of the
expansion from decelerating to accelerating is an intrinsic property of the
universe without need of an exotic dark energy. The cosmological constant
problem, the coincidence problem and the problem of phantom divide line
crossing are naturally solved.

\section{$\text{Field equations}$}

We start from the reduced action
\begin{equation}
S=\frac 1{2\kappa ^2}\int d\Omega \left( \sqrt{-g}f\left( \Gamma \right)
+L_m\right) ,
\end{equation}
where $\kappa ^2=8\pi G_N$ with the bare gravitational constant $G_N$,
\begin{equation}
\Gamma =\left( \left\{ _\sigma {}^\mu {}_\nu \right\} \left\{ _\rho {}^\nu
{}_\mu \right\} -\left\{ _\mu {}^\nu {}_\nu \right\} \left\{ _\sigma {}^\mu
{}_\rho \right\} \right) g^{\rho \sigma },
\end{equation}
is the reduced gravitational Lagrangian of general relativity \cite{10} with
the Christoffel symbol $\left\{ _\nu {}^\mu {}_\sigma \right\} $. $\Gamma $
is a quadratic form of the gravitational {\em strength } $\left\{ _\nu
{}^\mu {}_\sigma \right\} $ and identified with the kinetic energy of the
gravitational potential $g_{\mu \nu }$. The variational principle yields the
field equations for the metric $g_{\mu \nu }$:
\begin{equation}
f_\Gamma \left( R_{\rho \sigma }-\frac 12Rg_{\rho \sigma }\right) +\frac 12%
g_{\rho \sigma }\left( f_\Gamma \Gamma -f\left( \Gamma \right) \right)
-f_{\Gamma \Gamma }\Gamma _{,\mu }\frac{\partial \Gamma }{\partial g^{\rho
\sigma }{}_{,\mu }}=\kappa ^2T_{\rho \sigma }.
\end{equation}
where
\begin{equation}
f_\Gamma =\frac{\partial f}{\partial \Gamma },f_{\Gamma \Gamma }=\frac{%
\partial ^2f}{\partial \Gamma ^2},
\end{equation}
$R_{\rho \sigma }$ is the Ricci curvature tensor of $\left\{ _\sigma {}^\mu
{}_\nu \right\} $ and
\begin{equation}
T_{\rho \sigma }=-\frac 2{\sqrt{-g}}\frac{\delta L_m}{\delta g^{\rho \sigma }%
},
\end{equation}
is the energy-momentum of the matter fields. These equations can be
re-arranged in the Einstein-like form
\begin{equation}
\left( R_{\mu \nu }-\frac 12Rg_{\mu \nu }\right) =\frac 1{2f_\Gamma }\left(
f-f_\Gamma \Gamma \right) g_{\mu \nu }+\frac{f_{\Gamma \Gamma }}{f_\Gamma }%
\Gamma _{,\lambda }\frac{\partial \Gamma }{\partial g^{\mu \nu }{}_{,\lambda
}}+\frac{\kappa ^2}{f_\Gamma }T_{\mu \nu }.
\end{equation}

\section{$\text{Cosmological model}$}

We now investigate the cosmological dynamics for the models based on $%
f(\Gamma )$ gravity. In order to derive conditions for the cosmological
viability of $f(\Gamma )$ models we shall carry out a general analysis
without specifying the form of $f(\Gamma )$ at first. We consider a flat
Friedmann-Lemaitre-Robertson-Walker (FLRW) background with the metric
\begin{equation}
g_{\mu \nu }=\text{diag}\left( -1,a\left( t\right) ^2,a\left( t\right)
^2,a\left( t\right) ^2\right) ,
\end{equation}
where $a(t)$ is a scale factor. The non-vanishing components of the
Christoffel symbol are

\begin{eqnarray}
\left\{ _0{}^0{}_0\right\} &=&0,\left\{ _0{}^0{}_i\right\} =\left\{
_i{}^0{}_0\right\} =0,\left\{ _i{}^0{}_j\right\} =a\stackrel{\cdot }{a}%
\delta _{ij},  \nonumber \\
\left\{ _0{}^i{}_0\right\} &=&0,\left\{ _j{}^i{}_0\right\} =\left\{
_0{}^i{}_j\right\} =H\delta _j^i,\left\{ _j{}^i{}_k\right\}
=0,i,j,k,...=1,2,3.
\end{eqnarray}
and
\begin{equation}
\Gamma =\left( \left\{ _\sigma {}^\mu {}_\nu \right\} \left\{ _\rho {}^\nu
{}_\mu \right\} -\left\{ _\mu {}^\nu {}_\nu \right\} \left\{ _\sigma {}^\mu
{}_\rho \right\} \right) g^{\rho \sigma }=-6H^2.
\end{equation}
Here $H\equiv \stackrel{\cdot }{a}/a$ is the Hubble parameter and a dot
represents a derivative with respect to the cosmic time $t$. One can find
that $\Gamma $ is identified with the gravitational Lagrangian of the
cosmological model of $f(T)$ gravity given by Bengochea and Ferraro \cite{1}%
. The the field equations (6) take the forms
\begin{equation}
3H^2=-\frac 1{2f_\Gamma }\left( f+6H^2f_\Gamma \right) -18H^2\stackrel{\cdot
}{H}\frac{f_{\Gamma \Gamma }}{f_\Gamma }+\frac{\kappa ^2}{f_\Gamma }\rho ,
\end{equation}
\begin{equation}
-2\stackrel{\cdot }{H}-3H^2=\frac 1{2f_\Gamma }\left( f+6H^2f_\Gamma \right)
-18H^2\stackrel{\cdot }{H}\frac{f_{\Gamma \Gamma }}{f_\Gamma }+\frac{\kappa
^2}{f_\Gamma }p,
\end{equation}
which lead to
\begin{equation}
\stackrel{\cdot }{H}=-\frac{\kappa ^2}{2f_\Gamma }\left( \rho +p-\frac{36}{%
\kappa ^2}H^2\stackrel{\cdot }{H}f_{\Gamma \Gamma }\right) .
\end{equation}
It is easy to see that, in $f(\Gamma )$ gravity, the gravitational constant $%
\kappa ^2$ is replaced by an effective (time dependent) $\kappa _{\text{eff}%
}^2=\kappa ^2/f_\Gamma (\Gamma )$. On the other hand, it is reasonable to
assume that the present day value of $\kappa _{\text{eff}}^2$ is the same as
the $\kappa ^2$ so that we get the simple constraint :

\begin{equation}
\kappa _{\text{eff}}^2(z=0)=\kappa ^2\rightarrow f_\Gamma (\Gamma _0)=1,
\end{equation}
where $z$ is the redshift.

If $f\left( \Gamma \right) $ has the form
\begin{equation}
f\left( \Gamma \right) =\Gamma +\Phi \left( \Gamma \right) ,
\end{equation}
then the equations (10), (11) and (12) become
\begin{equation}
3H^2=\kappa ^2\left( \rho +\rho _{\text{de}}\right) ,
\end{equation}
\begin{equation}
-2\stackrel{\cdot }{H}-3H^2=\kappa ^2\left( p+p_{\text{de}}\right) ,
\end{equation}
and
\begin{equation}
\frac{\stackrel{\cdot \cdot }{a}}a=-\frac{\kappa ^2}6\left( \left( \rho
+\rho _{\text{de}}\right) +3\left( p+p_{\text{de}}\right) \right) ,
\end{equation}
where
\begin{equation}
\rho _{\text{de}}=-\frac 1{2\kappa ^2}\Phi -\frac 6{\kappa ^2}H^2\Phi
_\Gamma -\frac{18}{\kappa ^2}H^2\stackrel{\cdot }{H}\Phi _{\Gamma \Gamma },
\end{equation}
and
\begin{equation}
p_{\text{de}}=\frac 1{2\kappa ^2}\Phi +\frac 6{\kappa ^2}H^2\Phi _\Gamma +%
\frac 2{\kappa ^2}\stackrel{\cdot }{H}\left( \Phi _\Gamma -9H^2\Phi _{\Gamma
\Gamma }\right) ,
\end{equation}
are the density and the pressure of the ''dark energy'' $\Phi \left( \Gamma
\right) $, respectively. The state equation of the ''dark energy'' is then
\begin{equation}
w_{\text{de}}=\frac{p_{\text{de}}}{\rho _{\text{de}}}=-1-\frac{2\stackrel{%
\cdot }{H}\left( \Phi _\Gamma -18H^2\Phi _{\Gamma \Gamma }\right) }{\frac 12%
\Phi +6H^2\Phi _\Gamma +18H^2\stackrel{\cdot }{H}\Phi _{\Gamma \Gamma }}.
\end{equation}

Let $N=\ln a$. For any function $\Phi (a)$ we have $\Phi ^{\prime }=:\frac{%
d\Phi }{dN}=H\stackrel{\cdot }{\Phi }.$ The the equation (20) becomes
\begin{equation}
w_{\text{de}}==-1+\frac 13\frac{\Gamma ^{\prime }}\Gamma \frac{\Phi _\Gamma
+3\Gamma \Phi _{\Gamma \Gamma }}{\Phi /\Gamma -2\Phi _\Gamma +\frac 12\Gamma
^{\prime }\Phi _{\Gamma \Gamma }}.
\end{equation}
The equations (15) and (16) become
\begin{equation}
H^2=\frac{\kappa ^2}3\rho -\frac 16\Phi +\frac 13\Gamma \Phi _\Gamma +\Gamma
\stackrel{\cdot }{H}\Phi _{\Gamma \Gamma },
\end{equation}
\begin{equation}
\left( H^2\right) ^{\prime }=-\frac{2\kappa ^2p-\Gamma +\Phi -2\Gamma \Phi
_\Gamma }{3\Gamma \Phi _{\Gamma \Gamma }+2\Phi _\Gamma +2}.
\end{equation}
We see that a constant $\Phi $ acts just like a cosmological constant (dark
energy), and $\Phi $ linear in $\Gamma $ (i.e. $\Phi _\Gamma =$constant) is
simply a redefinition of gravitational constant $\kappa ^2$. The the
equation (23) can be written as
\begin{equation}
\frac 16\left( 1+\Phi _\Gamma -\frac 32\Gamma \Phi _{\Gamma \Gamma }\right)
\Gamma ^{\prime }=-\frac \Gamma 2+\frac 12\Phi -\Gamma \Phi _\Gamma +\kappa
^2p.
\end{equation}
Taking a universe with only dust matter so
\begin{equation}
p=0,
\end{equation}
we have
\begin{equation}
-\frac{1+\Phi _\Gamma -\frac 32\Gamma \Phi _{\Gamma \Gamma }}{3\Gamma \left(
1-\Phi /\Gamma +2\Phi _\Gamma \right) }d\Gamma =dN,
\end{equation}
and then we find the solution $\Gamma (a)$ in closed form:
\begin{equation}
a\left( \Gamma \right) =\exp \left\{ -\frac 13\int_{-6H_0^2}^\Gamma \frac{dx}%
x\frac{1+\Phi _x\left( x\right) -\frac 32x\Phi _{xx}\left( x\right) }{1-\Phi
\left( x\right) /x+2\Phi _x\left( x\right) }\right\} .
\end{equation}

The equations (22), (23) and (27) take the same forms as the ones for $f(T)$
gravity given by Linder \cite{2}. In a sense our model can be considered as
a metric formulation of the $f(T)$ model.

The equations (22) and (23) i.e.
\begin{equation}
3H^2=-\frac 12\Phi -6H^2\Phi _\Gamma -18H^2\stackrel{\cdot }{H}\Phi _{\Gamma
\Gamma }+\kappa ^2\rho .
\end{equation}
\begin{equation}
-2\stackrel{\cdot }{H}-3H^2=\frac 12\Phi +6H^2\Phi _\Gamma +2\stackrel{\cdot
}{H}\left( \Phi _\Gamma -9H^2\Phi _{\Gamma \Gamma }\right) +\kappa ^2p.
\end{equation}
yield
\begin{equation}
\stackrel{\cdot }{H}=-\frac{\kappa ^2}{2\left( 1+\Phi _\Gamma -18H^2\Phi
_{\Gamma \Gamma }\right) }\left( \rho +p\right) .
\end{equation}
So, $\frac{\kappa ^2}{1+\Phi _\Gamma -18H^2\Phi _{\Gamma \Gamma }}$ is
simply a redefinition of gravitational constant $\kappa ^2$.

In the vacuum (30) gives a {\bf universal} de\ Sitter solution
\begin{equation}
\stackrel{\cdot }{H}=0,
\end{equation}
for {\em any function} $\Phi \left( \Gamma \right) $. Then (20) gives
\begin{equation}
w_{\text{de}}=-1,
\end{equation}
which means that the function $\Phi \left( \Gamma \right) $ plays the role
of the cosmological constant or dark energy.

The equation (29) gives

\begin{equation}
\stackrel{\cdot }{H}=-\frac{\kappa ^2p+3H^2+\frac 12\Phi +6H^2\Phi _\Gamma }{%
2\left( 1+\Phi _\Gamma -9H^2\Phi _{\Gamma \Gamma }\right) }.
\end{equation}
Substituting (33) into (20) yields
\begin{equation}
w_{\text{de}}=-\frac{\Phi +6H^2\Phi _\Gamma +54H^4\Phi _{\Gamma \Gamma
}\allowbreak +2\kappa ^2\left( 9H^2\Phi _{\Gamma \Gamma }-\Phi _\Gamma
\right) p}{\left( 1+\Phi _\Gamma \right) \left( \Phi +12H^2\Phi _\Gamma
\right) -18H^2\Phi _{\Gamma \Gamma }\left( 12H^2\Phi _\Gamma +3H^2+\Phi
\right) -18H^2\Phi _{\Gamma \Gamma }\kappa ^2p}.
\end{equation}

In the case
\begin{equation}
\Phi \left( \Gamma \right) =\alpha \left( -\Gamma \right) ^n=\alpha
6^nH^{2n},
\end{equation}
(33) and (34) become
\begin{equation}
\stackrel{\cdot }{H}=-\frac{\kappa ^2p+3H^2+\left( \frac 12-n\right) \alpha
6^nH^{2n}}{2\left( 1+3n\left( 3n-5\right) \alpha 6^{n-2}H^{2n-2}\right) },
\end{equation}
and
\begin{equation}
w_{\text{de}}=-\frac{3\left( 3n-2\right) \left( n-1\right) H^2+n\left(
3n-1\right) \kappa ^2p}{-3\left( n+1\right) \left( 3n-2\right)
H^2+\allowbreak 6^nn\left( 2n-1\right) \left( 3n-2\right) \alpha
H^{2n}\allowbreak -3n\left( n-1\right) \kappa ^2p}.
\end{equation}

For dust matter
\[
p=0,
\]
we have
\begin{equation}
\stackrel{\cdot }{H}=-\frac{3H^2+\left( \frac 12-n\right) \alpha 6^nH^{2n}}{%
2\left( 1+3n\left( 3n-5\right) \alpha 6^{n-2}H^{2n-2}\right) },
\end{equation}
\begin{equation}
w_{\text{de}}=-\frac{3\left( 3n-2\right) \left( n-1\right) H^2}{-3\left(
n+1\right) \left( 3n-2\right) H^2+\allowbreak 6^nn\left( 2n-1\right) \left(
3n-2\right) \alpha H^{2n}\allowbreak }.
\end{equation}

When
\[
n=2,
\]
i.e. for the gravitational Lagrangian
\begin{equation}
L_g=\sqrt{-g}\left( \Gamma +\alpha \Gamma ^2\right) ,
\end{equation}
(39) gives
\begin{equation}
w_{\text{de}}=-\frac 1{3\left( 24\alpha H^2-1\right) }.
\end{equation}
Letting
\[
w_{\text{de}}=-1
\]
and
\[
H_0=74km/sec/Mpc\simeq 2.4\times 10^{-18}sec^{-1}
\]
one can compute
\begin{equation}
\alpha =1.\,0145\times 10^{-5}\left( km/sec/Mpc\right) ^{-2}=9.\,6451\times
10^{33}sec^2.
\end{equation}
Then when
\begin{equation}
w_{de}=-\frac{1}{3},
\end{equation}
we obtain
\begin{equation}
H=90.\,631km/sec/Mpc.
\end{equation}
Observations of type Ia supernovae at moderately large redshifts ($z\sim 0.5$
to $1$) have led to the conclusion that the Hubble expansion of the universe
is accelerating \cite{11}. This is consistent also with microwave background
measurements \cite{12}. According to the result of \cite{13}, $%
H=90.\,631km/sec/Mpc$ corresponds to
\[
z\sim 0.88,
\]
which is consistent  with the observations.

The equation (38) now becomes
\begin{equation}
\stackrel{\cdot }{H}=\frac{3H^2\left( 18\alpha H^2-1\right) }{2\left(
6\alpha H^2+1\right) }.
\end{equation}
Using the formula
\begin{equation}
H=\frac{\stackrel{\cdot }{a}}a=-\frac 1{1+z}\stackrel{\cdot }{z},
\end{equation}
(45) can be rewritten as
\begin{equation}
-\frac{2\left( 6\alpha H^2+1\right) dH}{3H\left( 18\alpha H^2-1\right) }=%
\frac{dz}{\left( 1+z\right) },
\end{equation}
and integrated
\begin{equation}
z=\left( \frac{H^2}{H_0^2}\right) ^{1/3}\left( \frac{18\alpha H_0^2-1}{%
18\alpha H^2-1}\right) ^{4/9}-1.
\end{equation}
This function is illustrated in Fig. 1, which is roughly consistent with the
observations \cite{13}.

\begin{figure}[ht]
  \includegraphics[width=6cm]{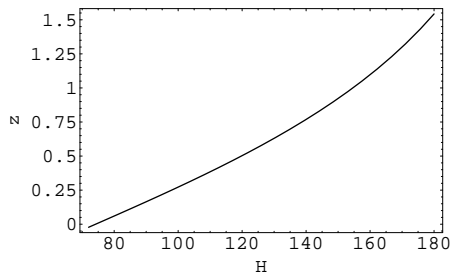}\\
  \caption{The function of $z=z(H)$.}\label{z-H}
\end{figure}

{\em \ }The equation (15) indicates that during the evolution of the
universe $H^2$ decreases owing to decreasing of the matter density $\rho $.
This makes $w_{\text{de}}$ descend during the evolution of the universe. The
evolution of the function $w_{\text{de}}=w_{\text{de}}\left( \alpha
H^2\right) $ given by (41) is illustrated in Fig. 2. According to (41) when $%
H^2=\frac 1{12\alpha }$, $w_{\text{de}}=$ $-\frac 13$, $\Phi \left( \Gamma
\right) $ changes from ''visible'' to dark as indicated by (17). If $H^2>%
\frac 1{12\alpha }$, it decelerates the expansion, if $H^2<\frac 1{12\alpha }
$, it accelerates the expansion. When $H^2=\frac 1{18\alpha }$, $w_{\text{de}%
}$ crosses the phantom divide line $-1$. In other words, the expansion of
the universe naturally includes a decelerating and an accelerating phase. $%
\alpha $ given by (42) can be seen as a new constant describing the
evolution of the universe.

\begin{figure}[ht]
  \includegraphics[width=6cm]{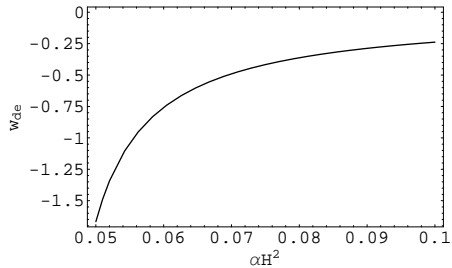}\\
  \caption{The evolution of $w_{de}$.}\label{wde}
\end{figure}

\end{document}